\documentclass{article}

\usepackage{arxiv}

\usepackage[utf8]{inputenc} 
\usepackage[T1]{fontenc}    
\usepackage{hyperref}       
\usepackage{url}            
\usepackage{booktabs}       
\usepackage{amsfonts}       
\usepackage{nicefrac}       
\usepackage{microtype}      
\usepackage{lipsum}
\usepackage{graphicx}
\usepackage{amsmath}
\usepackage{amssymb}
\usepackage{subcaption}
\usepackage{float}
\usepackage{caption}
\usepackage{booktabs}
\usepackage{array}
\usepackage{lipsum}
\usepackage{multirow} 
\usepackage[shortlabels]{enumitem}
\usepackage[numbers]{natbib}
\graphicspath{ {./images/} }

\title{Conceptual and Signal Processing Principles of
Spread-Spectrum Color Doppler Ultrasound}

\author{
 Kian Esmailian \\
  School of Biomedical Engineering \\
  and Robarts Research Institute \\
  Western University\\
  London, Ontario, Canada \\
  \texttt{kesmaili@uwo.ca} \\
   \And
 James C. Lacefield \\
  School of Biomedical Engineering, \\
  Department of Electrical \& Computer Engineering, \\
  Department of Medical Biophysics, \\
  and Robarts Research Institute\\
  Western University\\
  London, Ontario, Canada \\
  \texttt{jlacefie@uwo.ca} \\
}

\begin{document}
\maketitle
\begin{abstract}
The spread-spectrum color Doppler method provides an alternative to single-plane-wave color Doppler for ultrafast blood-flow imaging with a high Nyquist velocity, high frame rate, and long ensembles. Spread-spectrum images are acquired by transmitting plane waves at a sequence of azimuth angles that is repeated multiple times in a different random order during each repetition. Randomization of the plane-wave angle modulates in-focus and off-focus echoes in distinct manners that enable stationary clutter to be suppressed without compounding by application of a notching comb filter. In this paper, a simplified mathematical model for the slow-time signal acquired with a sequence of plane waves is used to explain the mechanics of the spread-spectrum method, investigate the effect of different discrete-time comb filter realizations on the accuracy of the resulting Doppler frequency estimates, compare the effectiveness of spread-spectrum to single-plane-wave imaging for spatial localization of a Doppler signal source, and characterize the method's sensitivity to background tissue motion.  The results establish a foundation to further improve the utility of the spread-spectrum method for practical applications involving non-stationary clutter and physiologically relevant flow patterns.
\end{abstract}

\keywords{Ultrafast ultrasound \and color Doppler \and plane-wave imaging \and spread-spectrum methods \and comb filters}

\section{Introduction}
The development of ultrafast color Doppler imaging has made possible many new applications of cardiovascular ultrasound \cite{villemain_ultrafast_2020,petrescu_concepts_2021}. In this paper, we focus on an ultrafast Doppler method for imaging high blood velocities at high frame rates, which is a valuable capability for scenarios such as estimating regurgitant volumes in patients with valvular heart disease \cite{buffle_evaluation_2025}.

Ultrafast images are acquired by transmitting planar or diverging waves to insonify the entire field of view simultaneously. A single unfocused transmission yields a low-resolution image (LRI). Most ultrafast Doppler methods employ coherent compounding of LRIs acquired at different plane-wave steering angles (or, equivalently, different virtual source positions for diverging waves) to reconstruct a high-resolution image that exhibits retrospective transmit focusing \cite{bercoff_ultrafast_2011}. One recognized disadvantage of coherent compounding for Doppler imaging is that it reduces the maximum blood velocity that can be estimated without aliasing because the slow-time sampling rate is reduced from the transmit pulse repetition frequency (PRF) to PRF/$M$ when LRIs are compounded at $M$ plane-wave angles. Even when no aliasing occurs, compounding produces a temporal low-pass filtering effect that can cause underestimation of systolic velocities \cite{ekroll_coherent_2015}.

Our laboratory previously introduced a spread-spectrum color Doppler method \cite{mansour_spread-spectrum_2016,manour_improved_2017} that is intended to avoid the disadvantages of coherent compounding while retaining some of the image quality benefits of using multiple plane-wave angles. The method’s key feature is a \textit{segmented} pulse sequence consisting of plane waves transmitted at $M$ unique azimuth angles that are repeated $L$ times. As is explained in greater detail in Sects.~\ref{sec:Slow-TimeMod} and ~\ref{sec:SS-Method} of this paper, the resulting slow-time signal from stationary clutter takes the form of a harmonic Fourier series that can be eliminated by a notching comb filter, while the signal power from the receive focus is spread approximately uniformly across the Doppler spectrum. This permits off-focus tissue echoes to be suppressed via digital filtering without the need for compounding, thereby enabling high blood velocities to be imaged accurately without aliasing. Experiments performed in straight-tube flow phantoms demonstrate that the spread-spectrum method performs well when the clutter scatterers are stationary \cite{mansour_spread-spectrum_2016,esmailian_background_2026}.

In our previous descriptions of spread-spectrum Doppler, essential aspects of the method were not clearly justified, which exposed the method to criticisms that it is unnecessarily complicated. In particular, in the spread-spectrum method, plane waves are transmitted in a randomized sequence and the resulting slow-time samples are shuffled into order of increasing angle prior to clutter filtering. The need for the randomization and shuffling steps was not demonstrated directly in prior publications. Furthermore, in the limiting case of one plane-wave angle, the spread-spectrum method simplifies to the single-plane-wave (SPW) methods demonstrated, for example, in \cite{osmanski_transthoracic_2014,guidi_real-time_2021}. An experimental comparison of spread-spectrum to SPW Doppler was presented in \cite{esmailian_background_2026}, but a conceptual justification for preferring the spread-spectrum method has not been elucidated. This comparison is important for justifying the additional complexity of the spread-spectrum method because it addresses our claim that the method retains some of the benefits of imaging at multiple plane-wave angles. This paper fills these knowledge gaps to lay a foundation for refining the method for more challenging experiments with non-stationary clutter and flow geometries more representative of Doppler echocardiography.

In the following sections, we introduce a conceptual model for noise-free slow-time signals acquired using spread-spectrum Doppler and apply that model to address the questions raised in the preceding paragraph. The model represents the Doppler signal source as a \textit{beacon} scatterer, \textit{i.e.}, a stationary point target whose reflectivity oscillates as a sinusoidal function of time. This approach permits the model to generate a slow-time signal with a narrowband Doppler frequency within a paradigm that invokes concepts such as analyzing an imaging system’s spatial impulse response. The intent is to present a model that prioritizes clarity of explanation over physical detail.

The beacon signal model is defined in Section~\ref{sec:Slow-TimeMod} and used to explain the steps of the spread-spectrum Doppler method in Section~\ref{sec:SS-Method}. Three methods for implementing a discrete-time notching comb filter are reviewed in Section~\ref{sec:CombFilt}. A series of simulations using the slow-time signal model are presented in Section~\ref{sec:Simulation}. The first simulation analyzes the effects of the comb filter realization and the sample shuffling steps on the accuracy of spread-spectrum estimates of the beacon frequency. The second simulation compares the abilities of spread-spectrum and SPW Doppler to localize a beacon signal. In the third simulation, the model is used to estimate the maximum background tissue velocity for which the spread-spectrum method can maintain its performance. Section~\ref{sec:Discussion} identifies observations from the beacon model that can inform further development of the spread-spectrum method and positions the method relative to other alias-resistant techniques developed for ultrafast Doppler.

\section{Slow-Time Signal Model}
\label{sec:Slow-TimeMod}

\begin{figure}[htbp]
    \centering

    \begin{subfigure}[t]{0.48\linewidth}
        \centering
        \includegraphics[width=\linewidth]{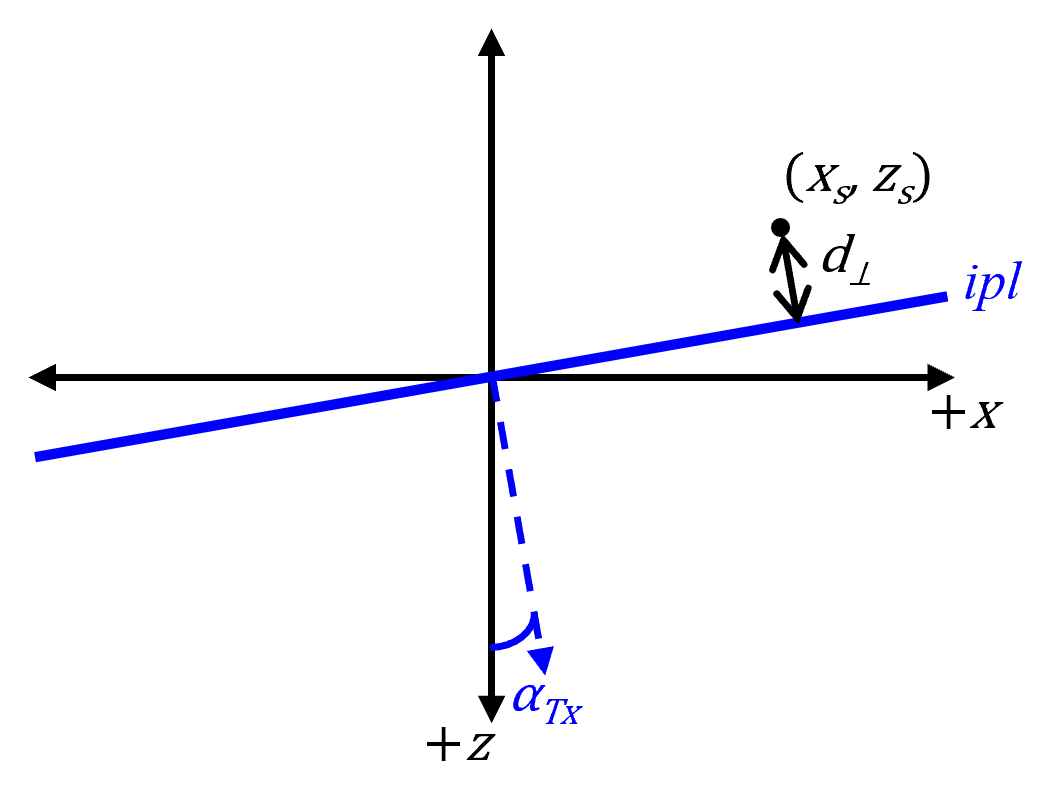}
        \caption{}
        \label{fig:Fig1a}
    \end{subfigure}
    \hfill
    \begin{subfigure}[t]{0.48\linewidth}
        \centering
        \includegraphics[width=\linewidth]{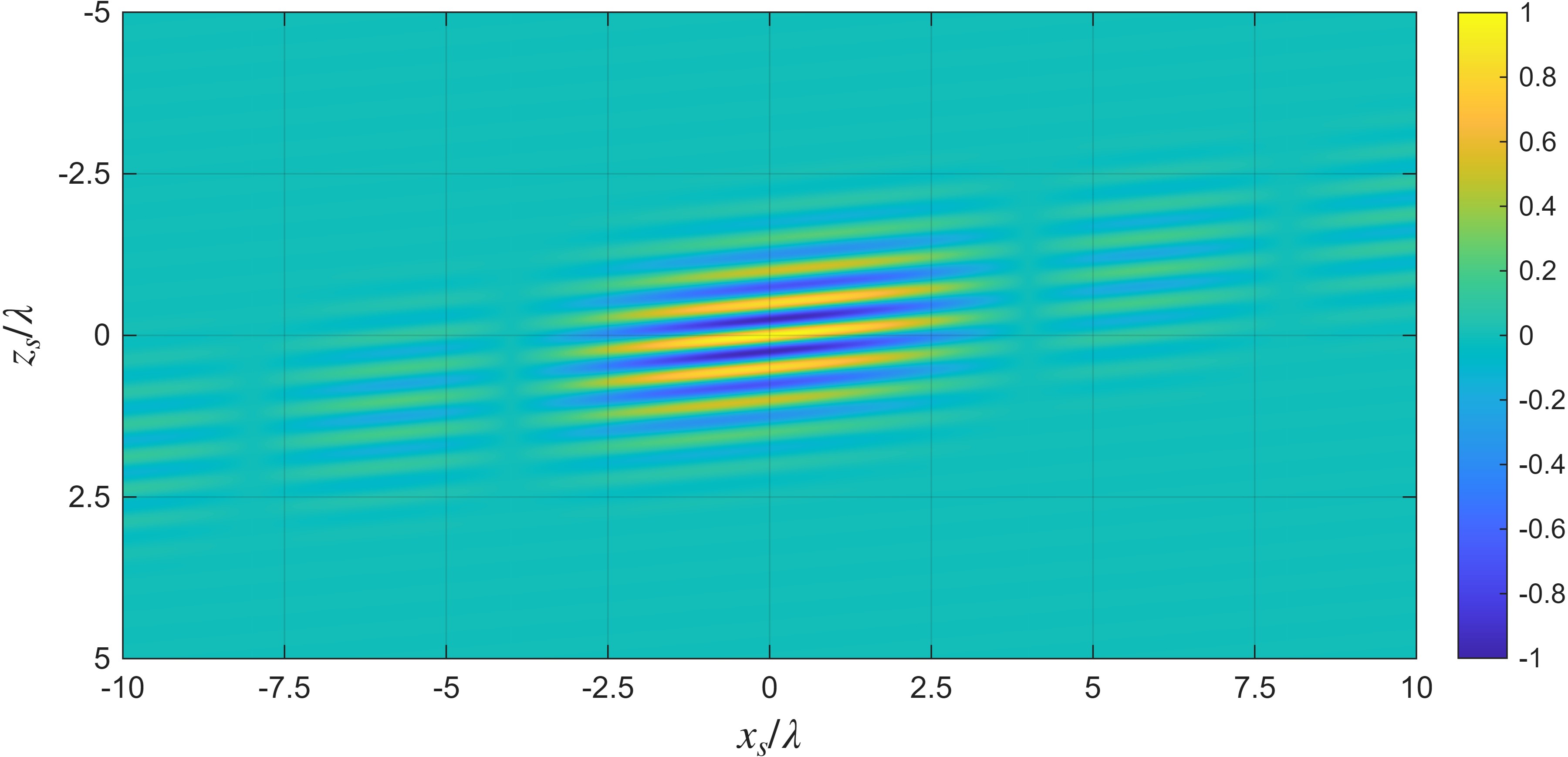}
        \caption{}
        \label{fig:Fig1b}
    \end{subfigure}

    \caption{(a) Coordinate system for the slow-time signal model. An incident planar wavefront transmitted at an azimuth angle $\alpha_{\mathrm{Tx}}$ is represented by an isophase line $(ipl)$ intersecting the receive focal point at the origin. The $x$ and $z$ axes correspond to the lateral and axial dimensions of the image, respectively. $d_{\perp}$ is the distance from a point scatterer located at $(x_s,z_s)$ to the isophase line of the plane wave. The blue dashed arrow indicates the propagation direction of the plane wave. (b) Normalized radio-frequency pulse-echo point-spread function computed using Eq.~\eqref{eq:slow-time-signal} with a plane-wave angle of $10^\circ$, full width at half-maximum pulse length equal to two wavelengths, and receiver $f$-number equal to 4.0.}
    \label{fig:Fig1}
\end{figure}

We seek a conceptual model for the slow-time signal acquired by transmitting a sequence of plane waves at a variable azimuth steering angle, $\alpha_{\mathrm{Tx}}$, and performing delay-and-sum receive beamforming at a single pixel location. We define a two-dimensional lateral $(x)$ and axial $(z)$ local coordinate system as shown in Fig.~\ref{fig:Fig1}(a), with its origin at the receive focus. The orientation of a plane wave can be described by the equation of an isophase line $(ipl)$ through the receive focus:
\begin{equation}
(\tan \alpha_{\mathrm{Tx}})x_{ipl} + z_{ipl} = 0.
\label{eq:isophase-line}
\end{equation}

Defining the isophase line to always pass through the receive focus represents the fact that, during receive beamformation, the fast-time signals acquired at different plane-wave angles are time-shifted to align the echoes from the receive focus. The signed distance from the isophase line to a scatterer at $(x_s,z_s)$ is:
\begin{equation}
d_{\perp}(x_s,z_s,\alpha_{\mathrm{Tx}})
=
\frac{(\tan \alpha_{\mathrm{Tx}})x_s + z_s}
{\sqrt{\tan^2\alpha_{\mathrm{Tx}} + 1}}.
\label{eq:signed-distance}
\end{equation}

The corresponding phase at $(x_s,z_s)$ of an incident plane wave with wavelength $\lambda$ is:
\begin{equation}
\phi_{pw}(x_s,z_s,\alpha_{\mathrm{Tx}})
=
-2\pi
\frac{d_{\perp}(x_s,z_s,\alpha_{\mathrm{Tx}})}
{\lambda},
\label{eq:plane-wave-phase}
\end{equation}
noting that points for which $d_{\perp} > 0$ correspond to a longer propagation distance than points along the isophase line and, hence, a time delay of the incident pulse. The reader will appreciate from Eqs.~\eqref{eq:isophase-line}--\eqref{eq:plane-wave-phase} that varying the plane wave's azimuth angle rotates the phase field about the origin, thereby modulating the echo received from an off-focus point target.

The theory section in the original publication of the spread-spectrum Doppler method \cite{mansour_spread-spectrum_2016} considered a narrowband signal model. We now extend the model to account for the effects of the plane wave's pulse envelope and receive beamforming on the magnitude of the slow-time signal. We assume a Gaussian pulse envelope with a full-width-at-half-maximum length $\ell_{\mathrm{FWHM}}$. The incident pulse magnitude at $(x_s,z_s)$ can be expressed using Eq.~\eqref{eq:signed-distance} as:
\begin{equation}
\left|P_{\mathrm{Tx}}(x_s,z_s,\alpha_{\mathrm{Tx}})\right|
=
\exp
\left(
\frac{-d_{\perp}^{2}(x_s,z_s,\alpha_{\mathrm{Tx}})}
{2\sigma^2}
\right),
\label{eq:pulse-envelope}
\end{equation}
where
\begin{equation}
\sigma =
\frac{\ell_{\mathrm{FWHM}}}
{2\sqrt{-2\log_e(0.5)}}
\end{equation}
is the scale parameter of the Gaussian envelope. For ease of intuition, we assume a receive aperture with rectangular apodization that is centered above its focal point, such that the receiver sensitivity pattern scales the magnitude of the pulse-echo signal from a point scatterer by approximately:
\begin{equation}
\left|S_{\mathrm{Rx}}(x_s)\right|
=
\left|
\operatorname{sinc}
\left(
\frac{x_s/\lambda}{F_{\mathrm{Rx}}/L_{\mathrm{Rx}}}
\right)
\right|,
\label{eq:receive-sensitivity}
\end{equation}
where $F_{\mathrm{Rx}}/L_{\mathrm{Rx}}$ is the $f$-number of the receive aperture. The transit time along the echo's return path contributes a phase shift that depends on the difference in range between the scatterer position and the receive focus at $(0,0)$:
\begin{equation}
\Delta r(x_s,z_s)
=
\sqrt{x_s^2 + (F_{\mathrm{Rx}}+z_s)^2}
-
F_{\mathrm{Rx}}.
\label{eq:range-difference}
\end{equation}

We simplify the model by invoking a paraxial approximation, $|x_s| \ll F_{\mathrm{Rx}}$, such that $\Delta r(x_s,z_s) \approx z_s$. Therefore, the phase shift due to the return propagation time is approximated by
\begin{equation}
\phi_{\mathrm{Rx}}(z_s)
\approx
-2\pi
\frac{z_s}{\lambda}.
\label{eq:receive-phase}
\end{equation}

Combining Eqs.~\eqref{eq:plane-wave-phase}--\eqref{eq:receive-sensitivity} and \eqref{eq:receive-phase}, the slow-time signal model for a unit-amplitude scatterer at $(x_s,z_s)$ becomes:
\begin{equation}
s(n)
=
\left|P_{\mathrm{Tx}}(x_s,z_s,\alpha_{\mathrm{Tx}})\right|
\left|S_{\mathrm{Rx}}(x_s)\right|
\operatorname{Re}
\left\{
\exp
\left[
j\phi_{pw}(x_s,z_s,\alpha_{\mathrm{Tx}})
+
j\phi_{\mathrm{Rx}}(z_s)
\right]
\right\}.
\label{eq:slow-time-signal}
\end{equation}

Figure~\ref{fig:Fig1}(b) shows the radio-frequency spatial impulse response estimated using Eq.~\eqref{eq:slow-time-signal} with $\ell_{\mathrm{FWHM}} = 2\lambda$, $F_{\mathrm{Rx}}/L_{\mathrm{Rx}} = 4$, and $\alpha_{\mathrm{Tx}} = 10^\circ$. This result can be viewed as an approximation to the point-spread functions for plane-wave imaging that were derived in \cite{alberti_mathematical_2017,chen_point_2019}.

\begin{figure}[htbp]
    \centering
    \includegraphics[width=0.96\linewidth]{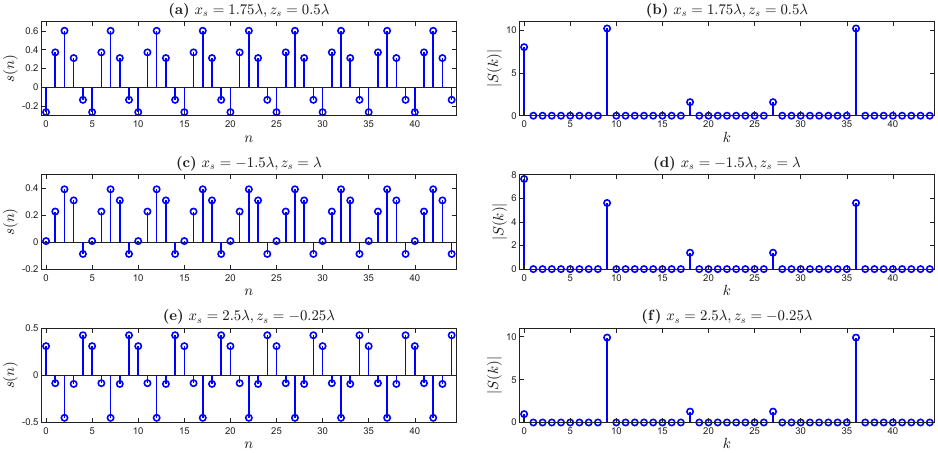}
    \caption{(a, c, e) Slow-time signals, $s(n)$, computed using Eq.~\eqref{eq:slow-time-signal} and (b, d, f) corresponding discrete magnitude spectra, $|S(k)|$, for single unit-amplitude scatterers at three different positions, $(x_s,z_s)$, relative to the receive focus. Each signal was synthesized with a $5~\text{angles} \times 9~\text{repetitions}$ plane-wave sequence with a maximum azimuth angle of $10^\circ$, a transmitted pulse length equal to two wavelengths, and a receiver $f$-number equal to 4.0.}
    \label{fig:Fig3}
\end{figure}

Consider a slow-time signal, $s(n)$, acquired from a single stationary point target by transmitting plane waves at $M=5$ uniformly spaced angles that are repeated in ascending order $L=9$ times. Let the transmit pulse length, receive $f$-number, and plane-wave angle sequence be the same as in those used in Fig.~\ref{fig:Fig1}(b). As illustrated for three different scatterer positions in Fig.~\ref{fig:Fig3}, Eq.~\eqref{eq:slow-time-signal} yields realizations of $s(n)$ that always consist of exactly $L$ cycles of a signal with a period of $M$ samples, such that its fundamental normalized frequency is $\hat{\omega}_0 = 2\pi/M$. However, the shape and magnitude of the periodic waveform can vary dramatically among different scatterer positions.

The discrete Fourier transform (DFT) of $s(n)$ yields a magnitude spectrum, $|S(k)|$, that consists of a non-zero component at $k=0$ if the signal average is not equal to zero, plus a U-shaped pattern of non-zero components at $k=L$ corresponding to the fundamental frequency, $\hat{\omega}_0$, and each of its harmonics up to $k=(M-1)L$; see Fig.~\ref{fig:Fig3}. This observation suggests that echoes from a stationary, off-focus scatterer can be removed from the slow-time signal by applying a notching comb filter whose stopband frequencies align with the non-zero components of $|S(k)|$. This filter can be thought of as a spread-spectrum extension of a conventional ``simple'' finite impulse response (FIR) stationary echo canceller \cite{jensen_stationary_1993}.

A stationary point target at the receive focus may be viewed as a limiting case of the first example. Substitution of $(x_s,z_s)=(0,0)$ into Eqs.~\eqref{eq:signed-distance}--\eqref{eq:receive-sensitivity} and \eqref{eq:receive-phase} results in $|P_{\mathrm{Tx}}|=1$, $|S_{\mathrm{Rx}}|=1$, $\phi_{pw}=0$, and $\phi_{\mathrm{Rx}}=0$ for all $\alpha_{\mathrm{Tx}}$, such that Eq.~\eqref{eq:slow-time-signal} yields a constant slow-time signal. Therefore, for a stationary target at the focus, $|S(k)|$ simplifies to a single non-zero component at $k=0$ that is also eliminated by the notching comb filter.

The model readily extends to a population of discrete stationary point scatterers that are randomly positioned near the focus. Under the assumption that the echoes from individual scatterers add linearly, the resulting slow-time signal and its discrete magnitude spectrum have the same characteristics that were described above for an isolated point scatterer. 

To enable the model to emulate a Doppler signal, one or more scatterers can be replaced by beacon targets whose reflectivity oscillates as a function of slow time. To simplify the following demonstrations, we specify the beacon's reflectivity to oscillate sinusoidally at a normalized frequency $\hat{\omega}_b = 2\pi f_b/\mathrm{PRF}$, where $f_b$ is the frequency of the beacon in hertz. The slow-time signal from an isolated beacon at the receive focus will track the oscillation in reflectivity.

\begin{figure}[htbp]
    \centering
    \includegraphics[width=0.96\linewidth]{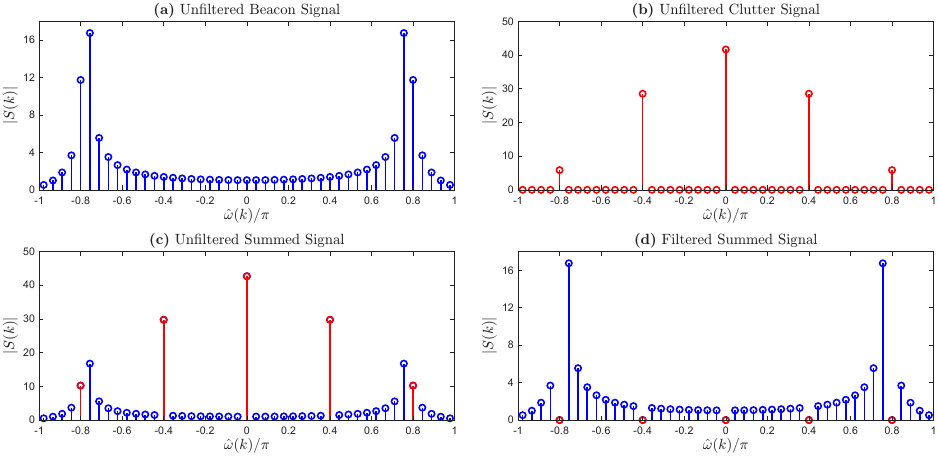}
    \caption{Illustrative example of a beacon signal superimposed on a stationary clutter background imaged using the same $5~\text{angles} \times 9~\text{repetitions}$ plane-wave sequence as in Fig.~\ref{fig:Fig3}. (a) Discrete magnitude spectrum, $|S(k)|$, for a $\hat{\omega}_b = 0.775\pi$ beacon at the receive focus plotted as a function of the normalized frequency of the spectral bins, $\hat{\omega}(k)$. (b) Discrete magnitude spectrum of the clutter signal. (c) Discrete magnitude spectrum of the coherent sum of the beacon and clutter signals with the harmonic components of the clutter signal highlighted in red. (d) Discrete magnitude spectrum of the summed signal after application of a notching comb filter to suppress the harmonic components of the clutter signal.}
    \label{fig:Fig5}
\end{figure}

Next, consider a single beacon at the receive focus that is surrounded by a population of stationary background scatterers and is insonified using the same $5~\text{angles} \times 9~\text{repetitions}$ sequence of plane waves used in the preceding examples. Let $\hat{\omega}_b$ take an arbitrary value that is less than $\pi$ but not necessarily equal to the sample frequencies of any of the $M \times L$ bins in $S(k)$. As shown in Fig.~\ref{fig:Fig5} for $\hat{\omega}_b=0.775\pi$, the discrete magnitude spectrum of the resulting slow-time signal, Fig.~\ref{fig:Fig5}(c), is the superposition of the spectrum of a windowed sinusoid from the beacon, Fig.~\ref{fig:Fig5}(a), with the harmonic Fourier series spectrum from the stationary scatterers, Fig.~\ref{fig:Fig5}(b). In this demonstration, we used an unrealistically high beacon-to-background ratio of approximately $-6$~dB to make the contributions of both signal components visually prominent in the magnitude spectra. Application of a notching comb filter nullifies the frequency bins in $|S(k)|$ that contain the harmonic components from the stationary scatterers and eliminates some of the signal components from the beacon, but enough of the beacon signal is retained that it is still feasible to obtain an accurate estimate of $\hat{\omega}_b$ from the filter's output signal, Fig.~\ref{fig:Fig5}(d). In this example, the beacon frequency estimate obtained via lag-one autocorrelation \cite{kasai_real-time_1985} was $\hat{\omega}_{est} = 0.751\pi$.

The worst-case scenario arises when $\hat{\omega}_b$ is exactly equal to one of the harmonic frequencies in the signal from the stationary scatterers. As an illustrative example, consider the same configuration as the previous example, except with $\hat{\omega}_b = 0.8\pi$. In this case, application of the notching comb filter completely eliminates the beacon signal as well as the stationary background echoes. Therefore, a different approach is needed to prevent the velocity estimator from having ``blind frequencies'', analogous to the blind velocities observed in moving target indicator radar \cite{shrader_clutter_2008}, at the harmonic frequencies of the clutter echoes.

\section{The Spread-Spectrum Doppler Method}
\label{sec:SS-Method}

A transmit pulse sequence for the spread-spectrum Doppler method, like the examples in Sect.~\ref{sec:Slow-TimeMod}, consists of plane waves transmitted at $M$ unique azimuth angles that are repeated $L$ times each. The first additional step in the method is to shuffle the order of the plane-wave angles so they are transmitted in a different pseudorandom order each time the set of angles is repeated. The transmit pulse sequence is thus a concatenation of $L$ segments. The first segment consists of $M$ plane waves whose azimuth angle varies in pseudorandom order, the second segment consists of the same $M$ plane-wave angles transmitted in a different pseudorandom order, and so on. We refer to the complete sequence of $M \times L$ plane waves as a \textit{segmented} pulse sequence and the resulting slow-time signal, whose samples are in the order they were acquired, as a \textit{time-ordered} signal.

\begin{figure}[htbp]
    \centering
    \includegraphics[width=0.96\linewidth]{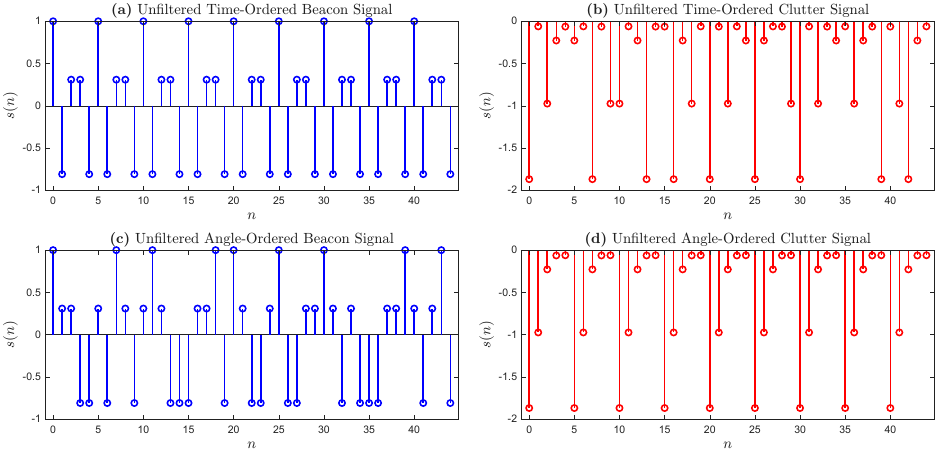}
    \caption{Effect of the sample-sorting step in the spread-spectrum Doppler method on unfiltered time-domain signal components imaged using a $5~\text{angles} \times 9~\text{repetitions}$ plane-wave sequence. (a) Time-ordered beacon signal with frequency $\hat{\omega}_b = 0.8\pi$. (b) Time-ordered stationary clutter signal. (c) Angle-ordered beacon signal. (d) Angle-ordered clutter signal.}
    \label{fig:Fig6}
\end{figure}

Figure~\ref{fig:Fig6} shows the components of the time-ordered signal for the blind-frequency scenario mentioned at the end of Sect.~\ref{sec:Slow-TimeMod}, $\hat{\omega}_b = 0.8\pi$, with randomization of the segmented plane-wave sequence. As one can infer from Eqs.~\eqref{eq:signed-distance} and \eqref{eq:plane-wave-phase}, randomly changing the value of $\alpha_{\mathrm{Tx}}$ produces a random phase modulation of the slow-time signal. This phase modulation causes the time-ordered signal from stationary off-focus scatterers, Fig.~\ref{fig:Fig6}(b), to become aperiodic. However, the time-ordered signal from a beacon at the receive focus, Fig.~\ref{fig:Fig6}(a), still tracks the sinusoidal fluctuation in the beacon reflectivity because $\phi_{pw}=\phi_{\mathrm{Rx}}=0$ for a scatterer at $(x_s,z_s)=(0,0)$ regardless of the value of $\alpha_{\mathrm{Tx}}$.

\begin{figure}[htbp]
    \centering
    \includegraphics[width=0.96\linewidth]{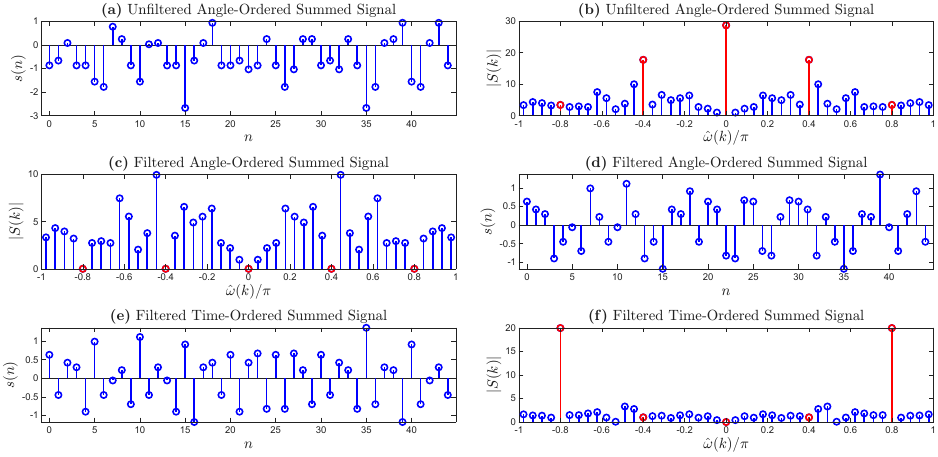}
    \caption{Slow-time signal processing steps in spread-spectrum color Doppler. (a) Time-domain angle-ordered signal, \textit{i.e.}, the coherent sum of the signals in Figs.~\ref{fig:Fig6}(c) and \ref{fig:Fig6}(d). (b) Discrete magnitude spectrum of the signal in panel (a), which is the input signal to the notching comb filter. The harmonic components of the angle-ordered clutter signal, Fig.~\ref{fig:Fig6}(d), are highlighted in red. (c) Discrete magnitude spectrum of the filter's angle-ordered output signal. The frequency bins that were nullified by the comb filter are highlighted in red. (d) Time-domain angle-ordered output signal. (e) Time-domain time-ordered output signal. (f) Discrete magnitude spectrum of the time-ordered output signal. The frequency components at the stopbands of the comb filter that were retained by the sample shuffling steps are highlighted in red. }
    \label{fig:Fig7}
\end{figure}

The second step of the spread-spectrum method is to sort the slow-time samples so they are in ascending order of $\alpha_{\mathrm{Tx}}$ within each segment of the sequence; we refer to the result as an \textit{angle-ordered} signal. This sorting restores the periodicity of the angle sequence and hence restores the periodicity of the slow-time signal from stationary off-focus scatterers, Fig.~\ref{fig:Fig6}(d). However, the sorting also disrupts the time alignment of the echoes from the receive focus, so the angle-ordered beacon signal, Fig.~\ref{fig:Fig6}(c), becomes aperiodic. The complete angle-ordered signal, which is the superposition of the angle-ordered beacon and background signals, is shown in Fig.~\ref{fig:Fig7}(a); its discrete magnitude spectrum is shown in Fig.~\ref{fig:Fig7}(b), with the harmonic frequencies of the stationary clutter signal in red and the other spectral components of the beacon signal in blue. Sorting the slow-time samples took a time-ordered beacon signal that was confined to the $\hat{\omega}_b=\pm 0.8\pi$ bins and dispersed its power across all frequency bins in the angle-ordered signal. This outcome is the basis for referring to the technique as a \textit{spread-spectrum} method; the name implies an analogy to the use of pseudorandom amplitude modulation to spread a narrowband information signal over a wide channel bandwidth in direct-sequence spread-spectrum communication systems \cite{haupt_spread_2020}.

Application of the notching comb filter nullifies the frequency bins at $\hat{\omega}_b=0$, $\pm 0.4\pi$, and $\pm 0.8\pi$, including the $\hat{\omega}_b=\pm 0.8\pi$ bins that contained all of the power of the original time-ordered beacon signal. The discrete magnitude spectrum and time-domain sequence of the resulting filtered angle-ordered signal are shown in Figs.~\ref{fig:Fig7}(c) and \ref{fig:Fig7}(d), respectively. The final step of the spread-spectrum method is to re-sort the filtered signal to return the samples to time order. Comparison of the filtered time-ordered signal, Fig.~\ref{fig:Fig7}(e), with the original time-ordered beacon signal, Fig.~\ref{fig:Fig6}(a), and inspection of the corresponding magnitude spectrum, Fig.~\ref{fig:Fig7}(f), illustrates the fact that nullification of $M$ out of $M \times L$ frequency bins by the notching comb filter distorts the beacon signal, but the $\hat{\omega}_b=\pm 0.8\pi$ bins remain the dominant spectral components of the filtered time-ordered signal. The beacon frequency estimate in this example was $\hat{\omega}_{est}=0.778\pi$.

Velocity estimation can be performed by applying any color Doppler velocity estimator to the filtered time-ordered signal. We employ lag-one autocorrelation \cite{kasai_real-time_1985} followed by the conventional display thresholding algorithms \cite{evans_ultrasonic_2011} in our implementation of the spread-spectrum Doppler method. A color Doppler image is constructed in a manner analogous to synthetic aperture beamformation by repeating the above analysis for every pixel in the Doppler field of view.

The key design choices when developing a plane-wave sequence for spread-spectrum Doppler are the number of unique plane-wave angles, $M$, and the number of repetitions, $L$, in the segmented sequence. The color Doppler image acquisition rate is $\mathrm{PRF}/(ML)$, so if ultrafast frame rates are desired, it is important to find the minimum product of $M$ and $L$ that yields acceptable velocity estimation accuracy. Exploration of this design space is beyond the scope of this paper, but, at this point in the analysis, the reader will appreciate that, since the notching comb filter nullifies $100M/(ML)\,\% = 100/L\,\%$ of the frequency bins in the angle-ordered signal, a low value of $L$ will more severely distort the signal from the receive focus and hence degrade the velocity estimation accuracy. However, the maximum unaliased Doppler frequency remains $\mathrm{PRF}/2$ for any combination of $M$ and $L$.

\section{Realization of the Notching Comb Filter}
\label{sec:CombFilt}

\subsection{Frequency-Sampling FIR Comb Filter}
\label{subsec:FreqSamplingFIR}

The demonstrations in Sects.~\ref{sec:Slow-TimeMod} and \ref{sec:SS-Method} employed the simplest discrete-time notching comb filter, which is the frequency-sampling FIR filter defined by a discrete frequency response that nullifies $M$ bins at the harmonics $k=0,L,2L,\ldots,(M-1)L$ corresponding to the normalized frequencies $\hat{\omega}=0,2\pi/M,2(2\pi/M),\ldots,(M-1)(2\pi/M)$:
\begin{equation}
H_{\mathrm{fsFIR}}(k)
=
\left[
\sum_{q=0}^{N-1} \delta(k-q)
\right]
-
\left[
\sum_{q=0}^{M-1} \delta(k-Lq)
\right],
\label{eq:HfsFIR-k}
\end{equation}
where $N=ML$ is the length of the pulse sequence. The inverse discrete Fourier transform of $H_{\mathrm{fsFIR}}(k)$ yields the filter's impulse response, which has a length of $M(L-1)+1$:
\begin{equation}
h_{\mathrm{fsFIR}}(n)
=
\frac{L-1}{L}\delta(n)
-
\frac{1}{L}
\sum_{q=1}^{L-1}
\delta(n-Mq).
\label{eq:hfsFIR-n}
\end{equation}

\begin{figure}[htbp]
    \centering
    \includegraphics[width=0.96\linewidth]{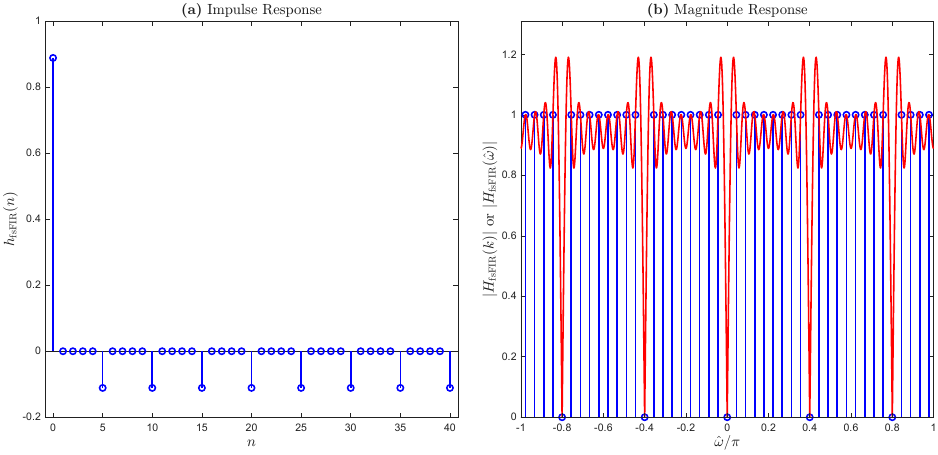}
    \caption{Frequency-sampling FIR implementation of a notching comb filter for a $5~\text{angles} \times 9~\text{repetitions}$ plane-wave sequence. (a) Impulse response, $h_{\mathrm{fsFIR}}(n)$. (b) Magnitude of the discrete frequency response, $|H_{\mathrm{fsFIR}}(k)|$, shown as a blue stem plot, and the discrete-time frequency response, $|H_{\mathrm{fsFIR}}(\hat{\omega})|$, shown as a red curve.}
    \label{fig:Fig8}
\end{figure}

As an illustrative example, $h_{\mathrm{fsFIR}}(n)$ for the $M=5$ angles, $L=9$ repetitions sequences considered in Sects.~\ref{sec:Slow-TimeMod} and \ref{sec:SS-Method} is shown in Fig.~\ref{fig:Fig8}(a). The filter's discrete-time frequency response, Fig.~\ref{fig:Fig8}(b), is readily obtained from Eq.~\eqref{eq:hfsFIR-n} as:
\begin{equation}
H_{\mathrm{fsFIR}}(\hat{\omega})
=
\frac{L-1}{L}
-
\frac{1}{L}
\sum_{q=1}^{L-1}
e^{-jMq\hat{\omega}}.
\label{eq:HfsFIR-omega}
\end{equation}

Increasing $L$ in Eq.~\eqref{eq:HfsFIR-omega} narrows the widths of the stopbands and alters the ripple in the passbands. The filter possesses a nonlinear phase response due to the asymmetry of its impulse response, but it has zero phase at each of its DFT bin frequencies.

The filter may be applied in the time domain by computing the convolution sum of $h_{\mathrm{fsFIR}}(n)$ with the angle-ordered slow-time signal. Alternatively, the filter may be applied in the frequency domain by computing the fast Fourier transform (FFT) of the angle-ordered signal, then multiplying the result by $H_{\mathrm{fsFIR}}(k)$. The latter approach implements a circular convolution. Circular convolution using a notching comb filter is an effective method to suppress an interfering signal that is integer periodic, \textit{i.e.}, its period is exactly an integer number of samples and the length of the signal is equal to an integer number of periods \cite{kuo_frequency-domain_1997}. If the clutter scatterers are stationary, the clutter signal in spread-spectrum Doppler will always have a period of $M$ samples and a length equal to $L$ periods. Therefore, circular convolution is an attractive approach in this context.

\subsection{Optimum Equiripple FIR Comb Filter}
\label{subsec:OptimumEquirippleFIR}

The optimum equiripple FIR notching comb filter introduced in \cite{zahradnik_design_2009} provides a second option that realizes a linear phase response at the expense of a modest increase in the complexity of computing the filter coefficients. The filter's generating function, {\cite{zahradnik_design_2009}, Eq.~(9)}, is a series summation of Chebyshev polynomials of the first kind from degree 0 to degree $\eta$. Reference \cite{zahradnik_design_2009} describes an efficient recursive algorithm for computing the filter coefficients that requires evaluation of only a small number of algebraic equations and hyperbolic trigonometric functions.

To obtain a specific realization of the optimum equiripple filter, the designer must specify three of the following four parameters: the degree of the generating function, $\eta$; the number of notches, $\rho$; the notch bandwidth; and the amplitude of the passband ripple. The values of $\eta$ and $\rho$ in combination determine the length of the filter's impulse response, $N_{\mathrm{equi}}=2 \eta \rho +1$. Given an $M$ angles $\times$ $L$ repetitions spread-spectrum pulse sequence, we have the design constraints $\rho=M$ and $N_{\mathrm{equi}}<ML$, where the latter constraint ensures the filter output will include samples that are valid linear convolution results.

In our implementation of this filter, we compute the largest integer $\eta$ that satisfies $N_{\mathrm{equi}}<ML$, then adjust the notch bandwidth to search for a filter realization that provides an acceptable balance of transition bandwidth and passband ripple. The impulse response of the optimal equiripple filter has $2 \eta +1$ non-zero coefficients. Through experimentation with the filter generating function, we discovered that it is possible to achieve an acceptable balance of transition bandwidth and passband ripple only if $\eta \geq 2$, so the impulse response will always contain at least five non-zero coefficients. This fact introduces a limitation that a satisfactory optimum equiripple filter will exist for our application only if $L \geq (4M+1)/M$.

\begin{figure}[htbp]
    \centering
    \includegraphics[width=0.96\linewidth]{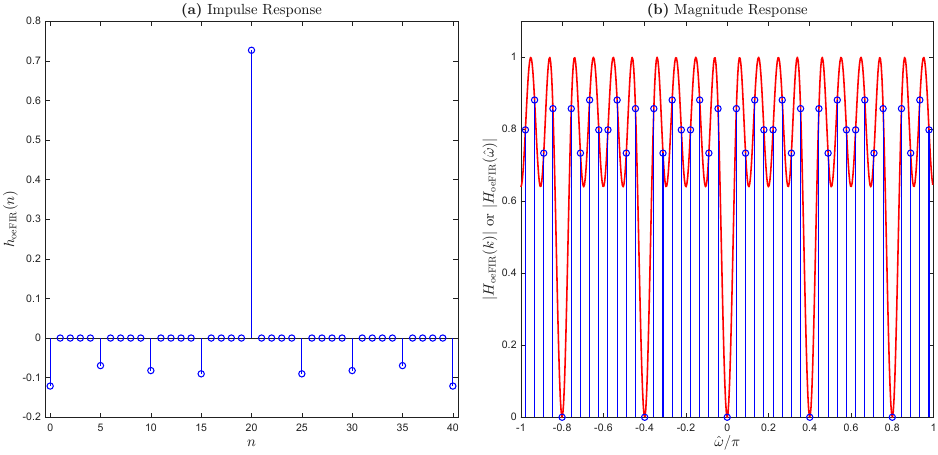}
    \caption{Optimum-equiripple FIR implementation of a notching comb filter for a $5~\text{angles} \times 9~\text{repetitions}$ plane-wave sequence. (a) Impulse response, $h_{\mathrm{oeFIR}}(n)$. (b) Magnitude of the discrete frequency response, $|H_{\mathrm{oeFIR}}(k)|$, shown as a blue stem plot, and the discrete-time frequency response, $|H_{\mathrm{oeFIR}}(\hat{\omega})|$, shown as a red curve.}
    \label{fig:Fig9}
\end{figure}

Figure~\ref{fig:Fig9} shows the impulse response, $h_{\mathrm{oeFIR}}(n)$, and frequency responses, $H_{\mathrm{oeFIR}}(k)$ and $H_{\mathrm{oeFIR}}(\hat{\omega})$, of an optimum equiripple notching comb filter for a 5 angles $\times$ 9 repetitions sequence, for which $\eta=4$ and the $-3$~dB notch bandwidth is $\pi/15$. Figure~\ref{fig:Fig9}(b) illustrates a disadvantage of the optimum equiripple filter if it is applied using circular convolution: $|H_{\mathrm{oeFIR}}(k)|$ possesses a non-uniform magnitude response across its passband frequency samples, whereas each of the passband samples of $|H_{\mathrm{fsFIR}}(k)|$ are unit amplitude.

\subsection{Infinite Impulse Response Comb Filter}
\label{subsec:IIRCombFilter}

A third option is an infinite impulse response (IIR) notching comb filter with $M$ notches at the same frequencies as the FIR filters described above, which is defined by the transfer function:
\begin{equation}
H_{\mathrm{IIR}}(z)
=
\frac{b_0-b_Mz^{-M}}
{1-a_Mz^{-M}},
\label{eq:HIIR-z}
\end{equation}
where $a_M$, $b_0$, and $b_M$ are positive scalars that have closed-form relationships to the widths of the stopbands and the signal attenuation at the notch frequencies \cite{orfanidis_comb_2023}. The IIR filter's frequency response,
\begin{equation}
H_{\mathrm{IIR}}(\hat{\omega})
=
\frac{b_0-b_Me^{-jM\hat{\omega}}}
{1-a_Me^{-jM\hat{\omega}}},
\label{eq:HIIR-omega}
\end{equation}
is free of ripple.

\begin{figure}[htbp]
    \centering
    \includegraphics[width=0.77\linewidth]{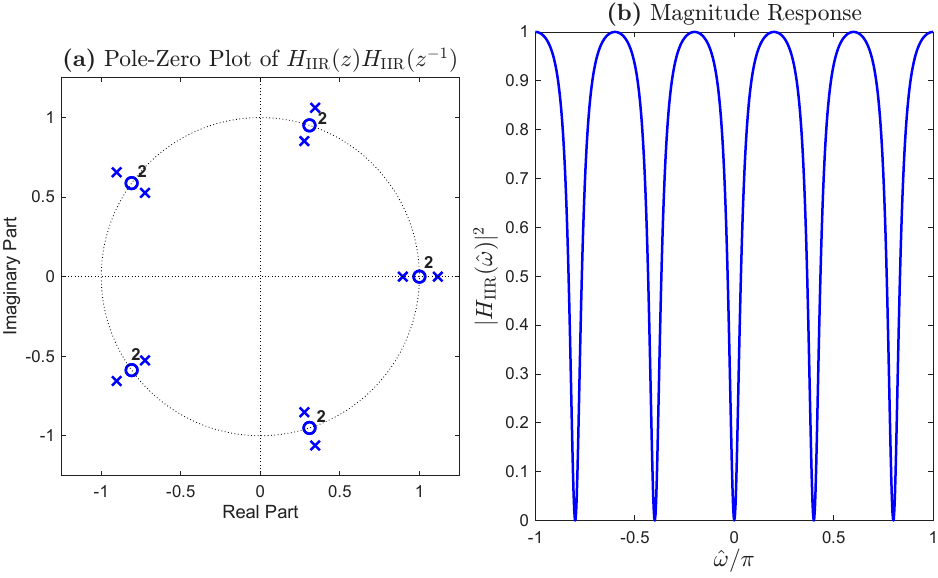}
    \caption{Zero-phase IIR implementation of a notching comb filter for a $5~\text{angles} \times 9~\text{repetitions}$ plane-wave sequence. (a) Pole-zero plot of the transfer function, $H_{\mathrm{IIR}}(z)H_{\mathrm{IIR}}(z^{-1})$. (b) Magnitude of the discrete-time frequency response, $|H_{\mathrm{IIR}}(\hat{\omega})|^2$.}
    \label{fig:Fig10}
\end{figure}

The phase response of $H_{\mathrm{IIR}}(\hat{\omega})$, like any IIR filter, is nonlinear. Zero-phase behavior can be achieved by forward-and-backward filtering, following the recommendation of Bjaerum \textit{et al.}~\cite{bjaerum_clutter_2002} for IIR clutter filtering in conventional focused-beam color Doppler. Figure~\ref{fig:Fig10} shows the pole-zero plot of the transfer function, $H_{\mathrm{IIR}}(z)H_{\mathrm{IIR}}(z^{-1})$, and the magnitude response of a zero-phase IIR notching comb filter, which can be obtained from Eq.~\eqref{eq:HIIR-omega} as $|H_{\mathrm{IIR}}(\hat{\omega})|^2$, with $M=5$ and $-3$~dB notch bandwidth equal to $\pi/15$.

\section{Simulations Using the Slow-Time Signal Model}
\label{sec:Simulation}

\subsection{Effect of Comb Filter Realization on Beacon Frequency Estimation}
\label{subsec:CombFilterFrequencyEstimation}

We considered a total of five options for realizing the notching comb filter: the frequency-sampling FIR filter or the optimum equiripple FIR filter employed with either circular or linear convolution, or the IIR filter employed with forward-and-backward filtering. We tested each of those filter realizations with and without pulse sequence randomization and shuffling between time and angle order. To select a filtering strategy, we prioritized identifying the approach that minimizes the distortion of an in-focus beacon signal by the filter. This analysis, perhaps surprisingly, recommends use of the frequency-sampling FIR filter with circular convolution. The following demonstration illustrates our rationale.

\paragraph{Simulation Method:}
All simulations presented in this paper are reported using normalized variables, \textit{i.e.}, position and pulse length are normalized by wavelength and slow time is normalized by pulse repetition interval (PRI~=~1/PRF). All simulations using multi-angle sequences employed $M=5$ plane-wave angles with $L=9$ repetitions of each angle. The plane-wave angles, $\alpha_{\mathrm{Tx}}$, were uniformly spaced from $-10^\circ$ to $10^\circ$ in increments of $5^\circ$. To construct a spread-spectrum sequence, an angle-ordered sequence was created, then the \texttt{randperm} function in MATLAB version 2024b (The MathWorks, Natick, MA) was used to generate a different random shuffling of the angles within each segment. For each plane-wave angle, the spatial impulse response to a scatterer or beacon at $(x_s,z_s)$ was estimated using Eqs.~\eqref{eq:isophase-line}--\eqref{eq:slow-time-signal} with $\ell_{\mathrm{FWHM}} = 2\lambda$ and $F_{\mathrm{Rx}}/L_{\mathrm{Rx}} = 4$.

The frequency-sampling FIR filter was implemented in the frequency domain using Eq.~\eqref{eq:HfsFIR-k} when circular convolution was employed or in the time domain using Eq.~\eqref{eq:hfsFIR-n} when linear convolution was employed. The impulse response of the optimal equiripple FIR filter was computed with degree $\eta=4$ and its $-3$~dB notch bandwidth set to $\pi/15$ using the algorithm detailed in \cite{zahradnik_design_2009} when linear convolution was employed. The impulse response of the optimal equiripple filter was zero-padded to length $ML$ and the FFT of the result was computed to obtain the frequency response for circular convolution. The coefficients of the IIR filter were computed using the \texttt{iircomb} function in the MATLAB Signal Processing Toolbox with its $-3$~dB notch bandwidth set to $\pi/15$. The IIR filter is applied using MATLAB's \texttt{filtfilt} function to implement forward-and-backward filtering; \texttt{filtfilt} employs the filter initialization method presented in \cite{gustafsson_determining_1996}. These steps yield the filters that are plotted in Figs.~\ref{fig:Fig8}--\ref{fig:Fig10}. 

When linear convolution is used, the output signal is cropped to retain the central samples that yield a filtered signal of the same length as the input signal so the shuffling from time to angle order, when it is applied, can be reversed after filtering. 

We isolated the effect of the pulse sequence and the notching comb filter on an in-focus beacon signal by evaluating the slow-time signal model for an isolated beacon at the origin with no other scatterers present. For each combination of pulse sequence and comb filter, the simulation was repeated while varying the normalized beacon frequency, $\hat{\omega}_b$, from $0$ to $\pi$ in steps of $\pi/(50ML)$. In each trial, the filtered signal power is computed in decibels with respect to the original signal power and the beacon frequency is estimated from the filtered, time-ordered signal using lag-one autocorrelation.

\begin{figure}[htbp]
    \centering
    \includegraphics[width=0.96\linewidth]{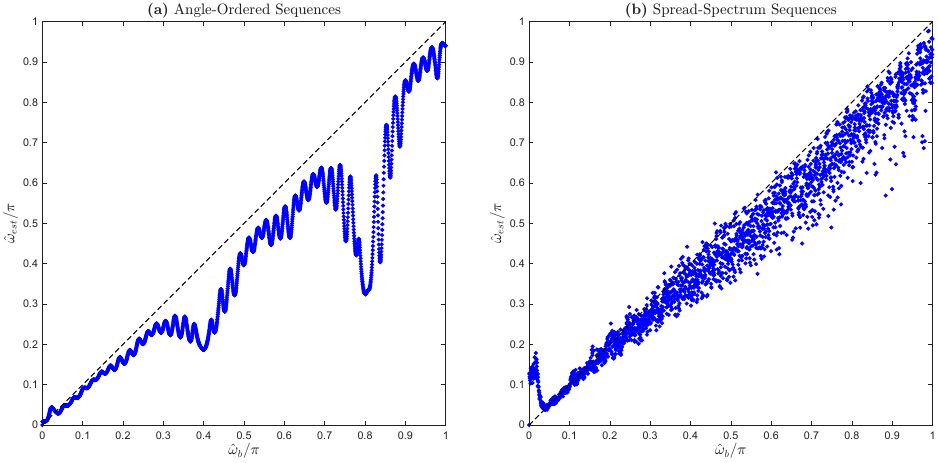}
    \caption{Beacon frequency estimation following application of an IIR notching comb filter to a $5~\text{angles} \times 9~\text{repetitions}$ plane-wave sequence. (a) Estimated frequency, $\hat{\omega}_{est}$, as a function of true beacon frequency, $\hat{\omega}_b$, using angle-ordered plane-wave sequences. (b) Frequency estimation using spread-spectrum plane-wave sequences. The black dashed line in each panel is the line of equality.}
    \label{fig:Fig11}
\end{figure}

\paragraph{Simulation Results:}
Figure~\ref{fig:Fig11} shows the estimated beacon frequency plotted against the true beacon frequency for the IIR filter when an angle-ordered sequence is used without the reshuffling steps, Fig.~\ref{fig:Fig11}(a), and when spread-spectrum sequences are used with the reshuffling steps, Fig.~\ref{fig:Fig11}(b). With the angle-ordered sequence, there is a consistent underestimation of the beacon frequency that is exacerbated at and near the filter's notch frequencies, $\hat{\omega}=0.4\pi$ and $0.8\pi$. With the spread-spectrum sequence, the scatter of the data points reflects the variability introduced by different randomizations of the plane-wave angles. Use of the spread-spectrum sequence avoids the local minima in estimated frequencies at the notch frequencies, but the underestimation of the beacon frequency persists, especially for $\hat{\omega}_b > 0.5\pi$, so the IIR filter was eliminated from further consideration on this basis.

\begin{figure}[htbp]
    \centering
    \includegraphics[width=0.96\linewidth]{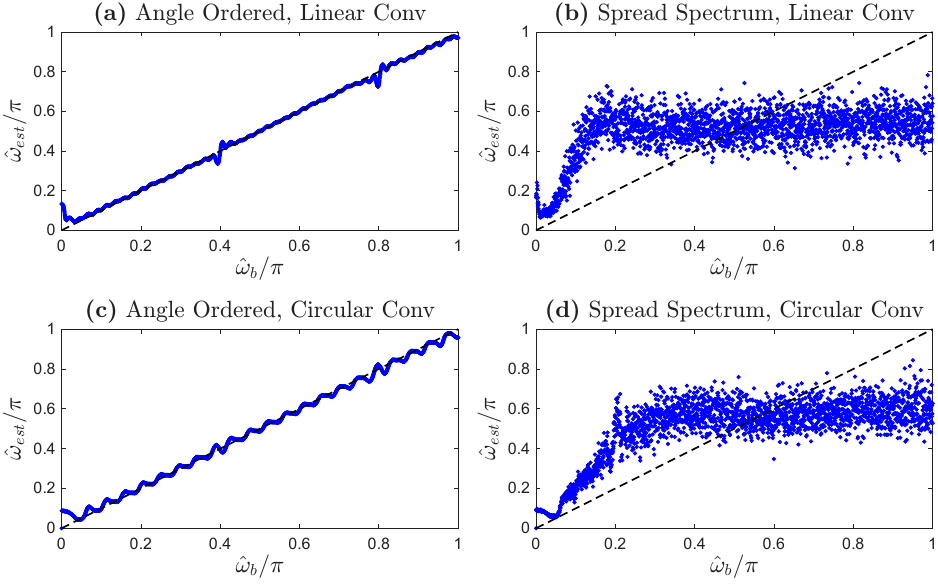}
    \caption{Beacon frequency estimation following application of an optimum equiripple FIR notching comb filter to a $5~\text{angles} \times 9~\text{repetitions}$ plane-wave sequence. (a) Estimated frequency, $\hat{\omega}_{est}$, as a function of true beacon frequency, $\hat{\omega}_b$, using angle-ordered plane-wave sequences and filtering via linear convolution. (b) Frequency estimation using spread-spectrum plane-wave sequences and filtering via linear convolution. (c) Frequency estimation using angle-ordered plane-wave sequences and filtering via circular convolution. (d) Frequency estimation using spread-spectrum plane-wave sequences and filtering via circular convolution. The black dashed line in each panel is the line of equality.}
    \label{fig:Fig12}
\end{figure}

Figure~\ref{fig:Fig12} shows the beacon frequency estimation performance using the optimum equiripple FIR filter with an angle-ordered sequence and linear convolution, spread-spectrum sequences and linear convolution, an angle-ordered sequence and circular convolution, and spread-spectrum sequences and circular convolution. The sample shuffling steps required with the spread-spectrum sequences distort the beacon signal sufficiently to prevent accurate frequency estimation, Figs.~\ref{fig:Fig12}(b) and \ref{fig:Fig12}(d). Accurate frequency estimates are obtained using either convolution method with the angle-ordered sequence, although the mean frequency estimation error is smaller with linear convolution, Fig.~\ref{fig:Fig12}(a), than with circular convolution, Fig.~\ref{fig:Fig12}(c).

\begin{figure}[htbp]
    \centering
    \includegraphics[width=0.96\linewidth]{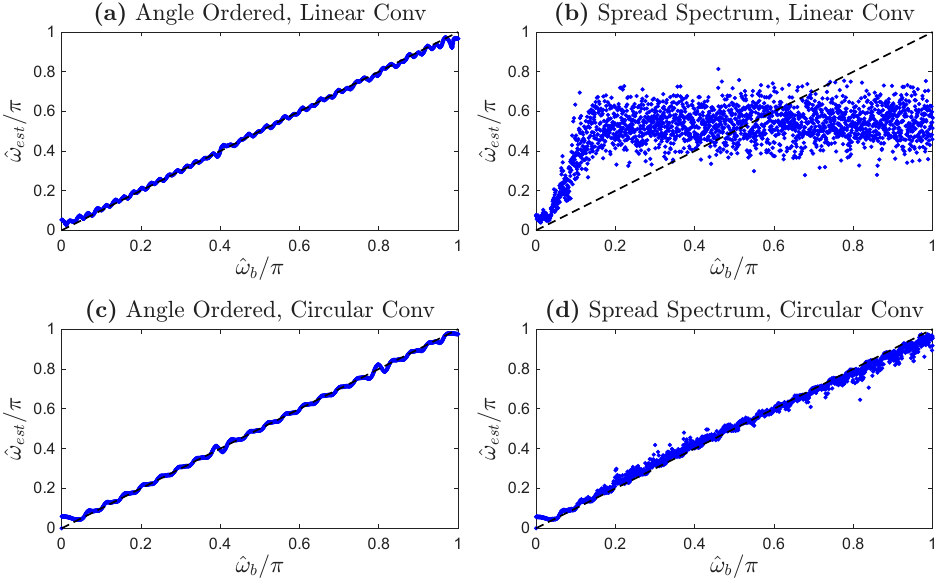}
    \caption{Beacon frequency estimation following application of a frequency-sampling FIR notching comb filter to a $5~\text{angles} \times 9~\text{repetitions}$ plane-wave sequence. (a) Estimated frequency, $\hat{\omega}_{est}$, as a function of true beacon frequency, $\hat{\omega}_b$, using angle-ordered plane-wave sequences and filtering via linear convolution. (b) Frequency estimation using spread-spectrum plane-wave sequences and filtering via linear convolution. (c) Frequency estimation using angle-ordered plane-wave sequences and filtering via circular convolution. (d) Frequency estimation using spread-spectrum plane-wave sequences and filtering via circular convolution. The black dashed line in each panel is the line of equality.}
    \label{fig:Fig13}
\end{figure}

Figure~\ref{fig:Fig13} shows the beacon frequency estimation performance using the frequency-sampling FIR filter in the same format as Fig.~\ref{fig:Fig12}. The accuracies of the frequency estimates using the angle-ordered sequence, Figs.~\ref{fig:Fig13}(a) and \ref{fig:Fig13}(c), and using spread-spectrum sequences with linear convolution, Fig.~\ref{fig:Fig13}(b), are similar to the results obtained using the optimum equiripple filter. However, the best frequency estimation accuracy with spread-spectrum sequences is obtained using the frequency-sampling FIR filter with circular convolution, Fig.~\ref{fig:Fig13}(d); compare to Figs.~\ref{fig:Fig11}(b), \ref{fig:Fig12}(b), \ref{fig:Fig12}(d), and \ref{fig:Fig13}(b).

If a multi-angle, non-compounding ultrafast color Doppler method was selected solely on the basis of the accuracy of the beacon frequency estimates, Figs.~\ref{fig:Fig11}--\ref{fig:Fig13} suggest that one should prefer an angle-ordered sequence and employ either FIR notching comb filter with linear convolution. However, recall that the motivation for the spread-spectrum method is to avoid blind velocities at the notch frequencies of the comb filter. Therefore, the attenuation of a beacon signal at a notch frequency should be minimal, especially since the blood-to-clutter ratio can be unfavorable in practical Doppler imaging. As shown in Table~\ref{tab:FilteredBeaconSignalPowers}, the combination of an angle-ordered sequence with circular convolution using either FIR filter yields the most severe signal loss at $\hat{\omega}_b=0.4\pi$ and $0.8\pi$, thereby eliminating those two options from consideration. On the other hand, the combination of a spread-spectrum sequence, the frequency-sampling FIR filter, and circular convolution retains the most beacon signal power at $\hat{\omega}_b=0.4\pi$ and $0.8\pi$.

\begin{table}[htbp]
    \centering
    \caption{Filtered beacon signal powers.}
    \label{tab:FilteredBeaconSignalPowers}
    \small
    \begin{tabular}{p{0.52\linewidth}
                    >{\centering\arraybackslash}p{0.11\linewidth}
                    >{\centering\arraybackslash}p{0.11\linewidth}
                    >{\centering\arraybackslash}p{0.11\linewidth}}
        \toprule
        \multirow{2}{*}{Plane-wave sequence and filtering method} &
        \multicolumn{3}{c}{Beacon signal attenuation at notch frequency [dB]} \\
        \cmidrule(lr){2-4}
        & $0$ & $0.4\pi$ & $0.8\pi$ \\
        \midrule
        Angle-ordered, optimum equiripple FIR, linear convolution & $-12.4$ & $-12.4$ & $-12.4$ \\
        Angle-ordered, optimum equiripple FIR, circular convolution & $-303$ & $-303$ & $-299$ \\
        Angle-ordered, frequency-sampling FIR, linear convolution & $-5.53$ & $-5.53$ & $-5.53$ \\
        Angle-ordered, frequency-sampling FIR, circular convolution & $-\infty$ & $-304$ & $-298$ \\
        Spread-spectrum, frequency-sampling FIR, circular convolution & $-\infty$ & $-0.26$ & $-0.12$ \\
        \bottomrule
    \end{tabular}
\end{table}

\begin{figure}[htbp]
    \centering
    \includegraphics[width=0.77\linewidth]{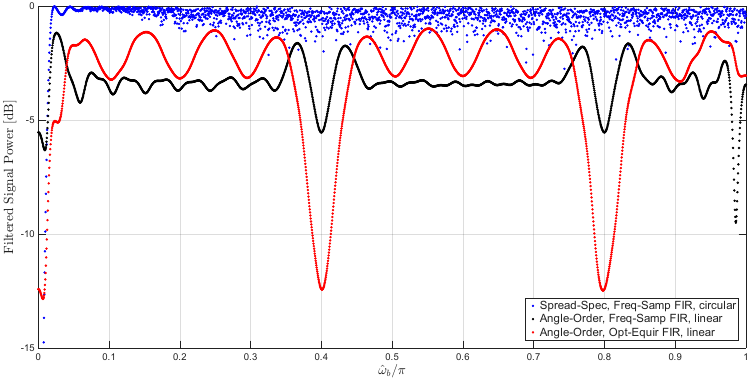}
    \caption{Filtered beacon signal power as a function of beacon frequency, $\hat{\omega}_b$, for the three FIR filtering options with the least loss of beacon signal. The blue data points were obtained using spread-spectrum sequences and the frequency-sampling FIR filter with circular convolution. The black data points were obtained using angle-ordered sequences and the frequency-sampling FIR filter with linear convolution. The red data points were obtained using angle-ordered sequences and the optimum-equiripple FIR filter with linear convolution. All results were computed using $5~\text{angles} \times 9~\text{repetitions}$ plane-wave sequences.}
    \label{fig:Fig14}
\end{figure}

To further explore the three strategies remaining under consideration, the filtered beacon signal powers are plotted as functions of beacon frequency in Fig.~\ref{fig:Fig14}, which shows that the filtered beacon power is greatest using the spread-spectrum sequence, the frequency-sampling FIR filter, and circular convolution at all frequencies greater than $0.0125\pi$. The strong attenuation of a beacon at $\hat{\omega}_b=0$ is also an advantage of the latter method because a $\hat{\omega}_b=0$ beacon is simply a stationary point scatterer and therefore should be eliminated by the clutter filter. The combination of the results in Fig.~\ref{fig:Fig13}(d), Fig.~\ref{fig:Fig14}, and Table~\ref{tab:FilteredBeaconSignalPowers} constitute our justification to continue development of the spread-spectrum method with the frequency-sampling FIR filter and circular convolution. The remaining simulations investigate only this implementation of the spread-spectrum method.

\subsection{Spatial Localization of Beacon Signals}
\label{subsec:ComparisonToSPWDoppler}

As stated in the Introduction, SPW Doppler can be viewed as a limiting case of the spread-spectrum method with $M=1$ angle. We have asserted that the additional complexity of the spread-spectrum method compared to SPW Doppler is justified by superior spatial localization of the blood signal. In this section, we demonstrate that the beacon signal model supports that hypothesis. We also compare SPW and spread-spectrum Doppler to plane-wave Doppler using coherent compounding as a positive control case. As a basis for fair comparison, we consider pulse sequences that yield identical frame rates.

\paragraph{Simulation Method:}
Spread-spectrum sequences with $5~\text{angles} \times 9~\text{repetitions}$ were constructed as described in Sect.~\ref{subsec:CombFilterFrequencyEstimation}. The single-plane-wave sequences consisted of 45 repetitions of a pulse transmitted at $\alpha_{\mathrm{Tx}}=0^\circ$. Coherent compounding simulations employed angle-ordered sequences of either three plane waves, $\alpha_{\mathrm{Tx}}=-10^\circ$, $0^\circ$, and $10^\circ$, repeated 15 times or five plane waves, $\alpha_{\mathrm{Tx}}$ identical to the spread-spectrum sequence, repeated 9 times. In the coherent compounding simulations, the slow-time signal computed for each segment of three or five plane waves was summed to yield a signal of length 15 or 9 samples, respectively.

As in Sect.~\ref{subsec:CombFilterFrequencyEstimation}, the slow-time signal model was evaluated for an isolated beacon, thereby simulating an idealized scenario where the clutter filter perfectly discriminates between the beacon and clutter signal components. Signal model computations were repeated many times while translating the beacon laterally from $x_s=-20\lambda$ to $20\lambda$ in steps of $\lambda/100$ while holding the axial position constant at $z_s=0$. For the SPW and coherent compounding simulations, no clutter filter was applied. For the spread-spectrum simulations, the frequency-sampling FIR filter was applied with circular convolution so the effects of the pulse sequence randomization and the comb filter on spatial localization of the beacon would be represented in the simulations.

For each imaging scenario, the power of the final slow-time signal was computed in decibels with respect to the signal power obtained with the beacon at the receive focus, $(x_s,z_s)=(0,0)$. Those signal powers were plotted as a function of the lateral position of the beacon to obtain a visualization of the sensitivity to the beacon analogous to a lateral beamplot in a beamforming study.

\begin{figure}[htbp]
    \centering
    \includegraphics[width=0.77\linewidth]{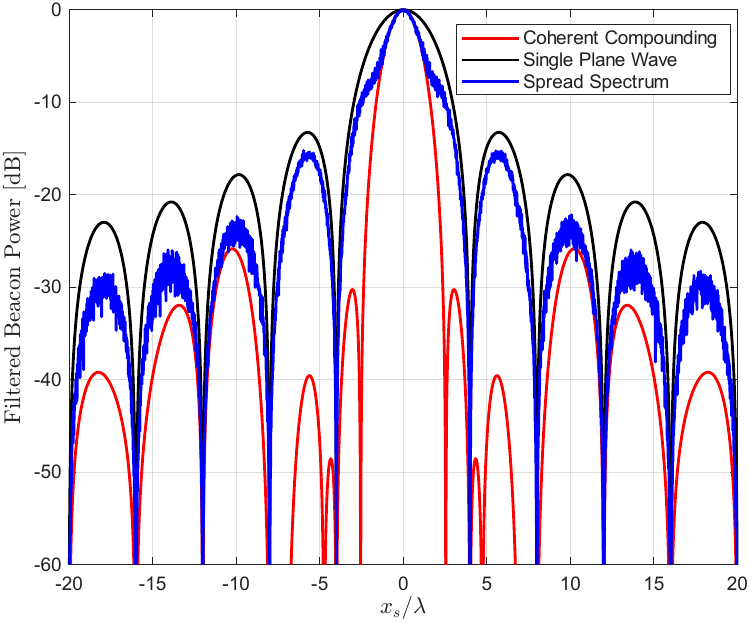}
    \caption{Filtered beacon signal power as a function of lateral beacon position, $x_s$, normalized by wavelength, $\lambda$. Computations were performed using a $5~\text{angles} \times 9~\text{repetitions}$ spread-spectrum sequence, shown in blue; a $5~\text{angles} \times 9~\text{repetitions}$ angle-ordered sequence with coherent compounding, shown in red; and a $1~\text{angle} \times 45~\text{repetitions}$ single-plane-wave sequence, shown in black.}
    \label{fig:Fig15}
\end{figure}

\paragraph{Simulation Results:}
The lateral sensitivity plots, Fig.~\ref{fig:Fig15}, are consistent with the expectation that the spread-spectrum method yields spatial localization of the beacon that is intermediate between ultrafast Doppler with coherent compounding and SPW Doppler. For visual clarity, only the 5-angle coherent compounding sequence is plotted in Fig.~\ref{fig:Fig15} because the two coherent compounding plots were almost overlapping. Observe that the main lobe of the spread-spectrum plot, blue, is narrower than the SPW main lobe, black, and is very close to the main-lobe width of the coherent compounding example, red, at signal powers $\geq -5$~dB. The sidelobe powers of the spread-spectrum data are also intermediate between the SPW and coherent compounding data. The high-spatial-frequency oscillations in the spread-spectrum plot reflect the variability introduced by using different pulse sequence randomizations at each beacon location.

Some key performance characteristics of the ultrafast Doppler methods are compared in Table~\ref{tab:PlaneWaveDopplerPerformance}. Although coherent compounding, as expected, provides the best spatial sensitivity to the beacon signal, the table emphasizes that SPW and spread-spectrum Doppler yield identical ensemble lengths and maximum unaliased beacon frequencies that are both substantially greater than the corresponding specifications for coherent compounding. The ensemble length is important because longer ensembles yield reduced variability in color Doppler frequency estimates \cite{guidi_real-time_2021,loupas_axial_1995}. The only difference between spread-spectrum and SPW Doppler in Table~\ref{tab:PlaneWaveDopplerPerformance} is the finer lateral sensitivity obtained in the beacon model simulations of the spread-spectrum method. The relatively coarse spatial sensitivity of SPW Doppler leads to bleeding artifacts in practical images \cite{ramalli_high-frame-rate_2020}. The results of this simulation are consistent with our flow-phantom results \cite{esmailian_background_2026} demonstrating that the spread-spectrum method substantially reduces those bleeding artifacts compared to SPW Doppler.

\begin{table}[htbp]
    \centering
    \caption{Performance characteristics of plane-wave Doppler methods with matching frame rates.}
    \label{tab:PlaneWaveDopplerPerformance}
    \small
    \begin{tabular}{p{0.37\linewidth}cccc}
        \toprule
        Plane-wave Doppler method and pulse sequence &
        FR &
        Ensemble length &
        Maximum unaliased $f_b$ &
        $-6$~dB lateral sensitivity \\
        \midrule
        Coherent compounding, $M \times L = 3 \times 15$ &
        $\mathrm{PRF}/45$ & $15$ & $\mathrm{PRF}/6$ & $2.20\lambda$ \\
        Coherent compounding, $M \times L = 5 \times 9$ &
        $\mathrm{PRF}/45$ & $9$ & $\mathrm{PRF}/10$ & $2.56\lambda$ \\
        Single plane wave, $M \times L = 1 \times 45$ &
        $\mathrm{PRF}/45$ & $45$ & $\mathrm{PRF}/2$ & $4.80\lambda$ \\
        Spread spectrum, $M \times L = 5 \times 9$ &
        $\mathrm{PRF}/45$ & $45$ & $\mathrm{PRF}/2$ & $3.03\lambda$ \\
        \bottomrule
    \end{tabular}

    \vspace{0.5em}
    \begin{minipage}{0.95\linewidth}
        \footnotesize
        Symbols: $M$ = number of unique plane-wave angles, $L$ = number of repeated transmissions of each plane wave, FR = Doppler frame rate, PRF = pulse repetition frequency, $f_b$ = beacon frequency, and $\lambda$ = wavelength.
    \end{minipage}
\end{table}

\subsection{Sensitivity to Tissue Motion}
\label{subsec:SensitivityToTissueMotion}

As illustrated in Sect.~\ref{sec:SS-Method}, the effectiveness of the notching comb filter for attenuating off-focus clutter relies on the integer periodic nature of the angle-ordered clutter signal, which in turn assumes that the background tissue is stationary so the same echo will be acquired from each plane-wave transmission at a given $\alpha_{\mathrm{Tx}}$. For modest levels of tissue motion, the angle-ordered clutter signal can be viewed as a superposition of a stationary signal component plus a disturbance due to tissue motion. For cases where the power of the disturbance component is low compared to the power of the stationary component, the notching comb filter will substantially attenuate the clutter signal and it will still be possible to construct an accurate spread-spectrum color Doppler image. In this section, we use the slow-time signal model to determine the extent of tissue velocities for which the spread-spectrum method will perform effectively under idealized conditions.

\paragraph{Simulation Method:}
Spread-spectrum sequences with $5~\text{angles} \times 9~\text{repetitions}$ were constructed as described in Sects.~\ref{subsec:CombFilterFrequencyEstimation} and \ref{subsec:ComparisonToSPWDoppler}, and the resulting slow-time signals were filtered using the frequency-sampling FIR filter with circular convolution. A population of 250 point scatterers was randomly distributed within a rectangular region of interest centered at the receive focus at $(x_s,z_s)=(0,0)$ and extending from $x_s=-6\lambda$ to $6\lambda$ laterally and from $z_s=-4\lambda$ to $4\lambda$ axially. The reflectivity of each scatterer was randomly assigned as a zero-mean, unit-variance Gaussian-distributed random variable.

Between each computation of a slow-time sample, the scatterers were translated at a constant velocity in the positive axial direction, \textit{i.e.}, away from the imaging aperture. Signal model computations were repeated while increasing the axial velocity of the scatterers from $10^{-5}\lambda/\mathrm{PRI}$ to $0.025\lambda/\mathrm{PRI}$ in 1000 logarithmically spaced increments. This yielded a maximum scatterer displacement of $44 \times 0.025\lambda = 1.1\lambda$ over the duration of the $5 \times 9$ spread-spectrum sequence. The initial scatterer positions were distributed to ensure that at least 50 scatterers would be located within the main lobe of the point-spread function illustrated in Fig.~\ref{fig:Fig1}(b) for each time step of a simulation.

At each tissue velocity, the filtered signal power was computed in decibels with respect to the unfiltered signal power. The simulation was repeated twice: once while generating new pulse sequence randomizations and initial scatterer positions at each tissue velocity to demonstrate the variability of the results and a second time using the same pulse sequence and initial scatterer positions at all tissue velocities to isolate the effect of tissue velocity on the power of the filtered clutter signal.

\begin{figure}[htbp]
    \centering
    \includegraphics[width=0.96\linewidth]{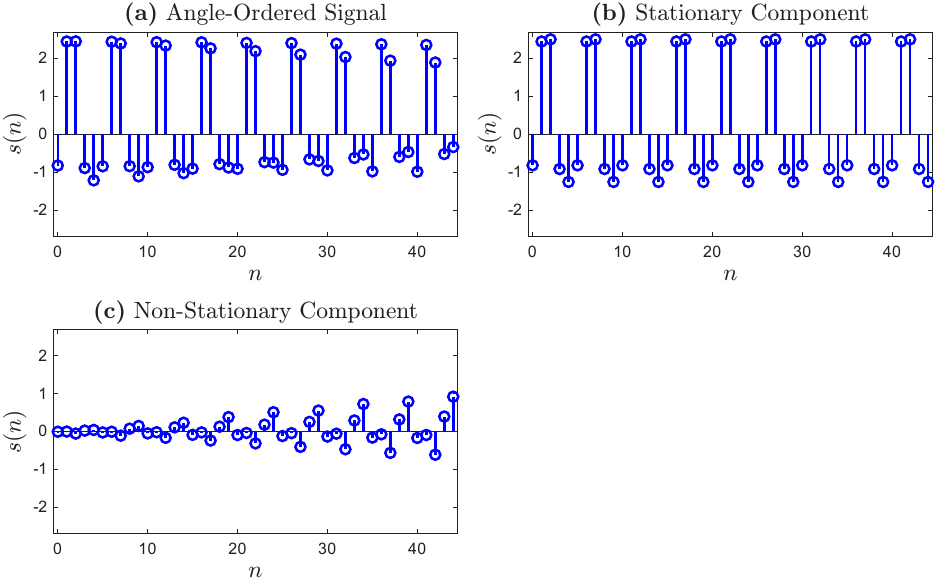}
    \caption{Illustrative example of (a) an unfiltered angle-ordered clutter signal in the presence of slow tissue motion, which can be considered a superposition of (b) a stationary clutter signal and (c) a growing disturbance due to the cumulative displacement of the scatterers. Computations were performed using a $5~\text{angles} \times 9~\text{repetitions}$ spread-spectrum sequence.}
    \label{fig:Fig16}
\end{figure}

\paragraph{Simulation Results:}
Figure~\ref{fig:Fig16} illustrates the decomposition of an angle-ordered unfiltered signal, Fig.~\ref{fig:Fig16}(a), into a stationary, integer periodic component, Fig.~\ref{fig:Fig16}(b), plus a disturbance due to tissue motion, Fig.~\ref{fig:Fig16}(c). In this example, the power of the filtered clutter signal was approximately $-20$~dB with respect to the unfiltered signal. Figure~\ref{fig:Fig16}(c) demonstrates that the envelope of the disturbance component increases monotonically as a function of slow time when the tissue velocity is constant.

\begin{figure}[htbp]
    \centering
    \includegraphics[width=0.77\linewidth]{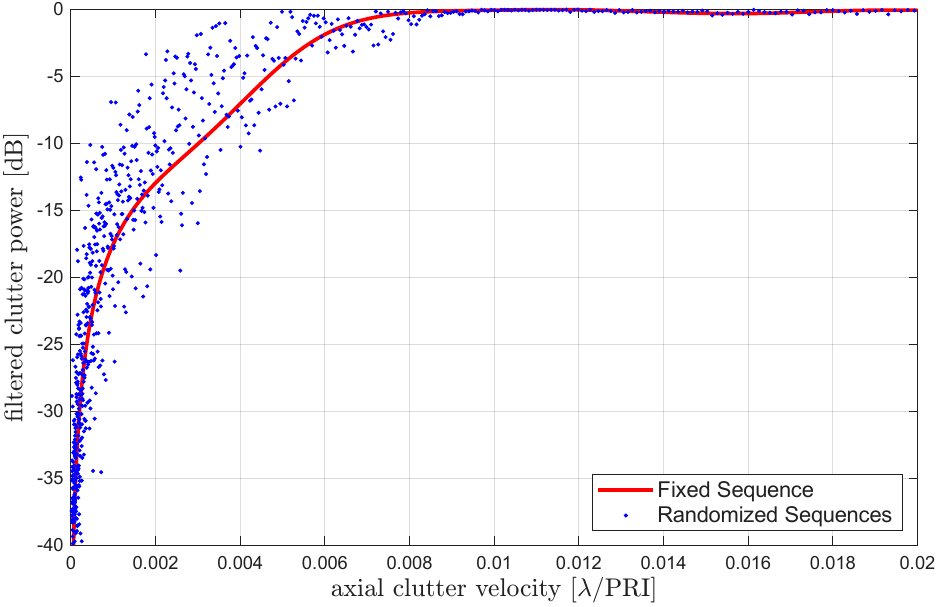}
    \caption{Filtered clutter signal power as a function of normalized axial tissue velocity in units of wavelengths, $\lambda$, per pulse repetition interval, PRI. The blue dots depict the results of computations performed using different plane-wave sequences and different initial scatterer positions at each velocity. The red curve shows a representative series of computations performed using an identical plane-wave sequence and the same initial scatterer positions at each velocity. All computations were performed using $5~\text{angles} \times 9~\text{repetitions}$ spread-spectrum sequences.}
    \label{fig:Fig17}
\end{figure}

Figure~\ref{fig:Fig17} shows the filtered clutter power as a function of axial tissue velocity for the simulation in which different pulse sequence randomizations and initial scatterer positions were used at each velocity, blue dots, and a representative simulation in which the same pulse sequence and initial scatterer positions were used at all velocities, red curve. The lower bound on the tissue velocity the spread-spectrum method can reliably tolerate can be read from Fig.~\ref{fig:Fig17} as the highest velocity for which the filtered clutter power never exceeds a given threshold. The upper bound on the range of tissue velocities the spread-spectrum method inconsistently tolerates can be defined by the maximum velocity for which the filtered clutter power remains below the threshold in at least one trial. Table~\ref{tab:TissueVelocityThresholds} lists these results for thresholds ranging from $-40$ to $-10$~dB, reported as normalized velocities and as the corresponding physical velocities for the specific scenario of a 5~MHz plane-wave center frequency and a 10~kHz PRF.

\begin{table}[htbp]
    \centering
    \caption{Axial tissue velocities yielding selected filtered clutter signal powers.}
    \label{tab:TissueVelocityThresholds}
    \small
    \begin{tabular}{ccccc}
        \toprule
        \multirow{2}{*}{Power threshold} &
        \multicolumn{2}{c}{Reliably tolerated velocity} &
        \multicolumn{2}{c}{Maximum tolerated velocity} \\
        \cmidrule(lr){2-3}
        \cmidrule(lr){4-5}
        & Normalized & \begin{tabular}{@{}c@{}}$f_0=5$~MHz,\\ $\mathrm{PRF}=10$~kHz\end{tabular}
        & Normalized & \begin{tabular}{@{}c@{}}$f_0=5$~MHz,\\ $\mathrm{PRF}=10$~kHz\end{tabular} \\
        \midrule
        $-40$~dB & $1.60 \times 10^{-5}\lambda/\mathrm{PRI}$ & $0.0493$~mm/s & $2.42 \times 10^{-4}\lambda/\mathrm{PRI}$ & $0.744$~mm/s \\
        $-30$~dB & $2.62 \times 10^{-5}\lambda/\mathrm{PRI}$ & $0.0806$~mm/s & $7.18 \times 10^{-4}\lambda/\mathrm{PRI}$ & $2.21$~mm/s \\
        $-20$~dB & $1.47 \times 10^{-4}\lambda/\mathrm{PRI}$ & $0.454$~mm/s & $1.81 \times 10^{-3}\lambda/\mathrm{PRI}$ & $5.56$~mm/s \\
        $-10$~dB & $9.51 \times 10^{-4}\lambda/\mathrm{PRI}$ & $2.93$~mm/s & $4.48 \times 10^{-3}\lambda/\mathrm{PRI}$ & $13.8$~mm/s \\
        \bottomrule
    \end{tabular}

    \vspace{0.5em}
    \begin{minipage}{0.95\linewidth}
        \footnotesize
        Symbols: $f_0$ = plane-wave center frequency, PRF = pulse repetition frequency, $\lambda$ = wavelength, and PRI = pulse repetition interval.
    \end{minipage}
\end{table}

\section{Discussion}
\label{sec:Discussion}

The primary observations from the slow-time signal model and these simulations are:

\begin{enumerate}[(1)]
\item The spread-spectrum Doppler method provides identically high unaliased blood velocities and long ensemble lengths as single-plane-wave Doppler while also yielding more accurate spatial localization of the blood signal than SPW Doppler (Fig.~\ref{fig:Fig15} and Table~\ref{tab:PlaneWaveDopplerPerformance}).

\item The additional complexity introduced by pulse sequence randomization and shuffling from time order to angle order and back to time order is justified by the elimination of blind velocities (Figs.~\ref{fig:Fig6}--\ref{fig:Fig7}).

\item Blind velocities are most effectively avoided if the notching comb filter is implemented as a frequency-sampling FIR filter and applied using circular convolution (Fig.~\ref{fig:Fig14} and Table~\ref{tab:FilteredBeaconSignalPowers}).

\item The spread-spectrum method does not require perfectly stationary background tissue to function effectively, but the range of tissue velocities over which it maintains its performance is limited in its current implementation (Fig.~\ref{fig:Fig17} and Table~\ref{tab:TissueVelocityThresholds}).
\end{enumerate}

The spread-spectrum Doppler method is an alternative to SPW Doppler for applications that would benefit from a capability to accurately image high, rapidly changing blood velocities. These scenarios often arise in ultrafast Doppler echocardiography, e.g., \cite{buffle_evaluation_2025,osmanski_transthoracic_2014}, where diverging waves are typically employed rather than plane waves to compensate for the small apertures of phased array transducers. Although we have not tested the spread-spectrum method with diverging wave transmissions, the technique should be readily adaptable to that approach. The transition from planar to diverging waves alters the footprint of the point-spread function, but the fundamental principle that a stationary tissue background will yield an integer-periodic clutter signal will still apply.

All simulations in this paper used the same spread-spectrum sequence parameters. The 45-pulse sequence length was chosen to represent imaging at a high Doppler frame rate; e.g., a $\mathrm{PRF}/45$ frame rate, Table~\ref{tab:PlaneWaveDopplerPerformance}, equates to 222 frames/second at a 10~kHz PRF. However, it is possible that the performance of the spread-spectrum method could be further improved by experimenting with different combinations of the number of plane waves, number of repetitions, and range of plane-wave angles, subject to a constraint that the sequence length achieves an acceptable frame rate. An optimization search in this manner is beyond the scope of this paper.

The simulations in Sect.~\ref{subsec:SensitivityToTissueMotion} confirm that the spread-spectrum method cannot tolerate substantial tissue motion, which is a concern for a technique intended for Doppler echocardiography. However, this concern may be addressed by replacing the Fourier-domain comb filter with a singular value decomposition (SVD) clutter filter as demonstrated for ultrafast Doppler echocardiography in \cite{papadacci_4d_2019}. The long ensemble lengths of the spread-spectrum method compared to methods employing coherent compounding, Table~\ref{tab:PlaneWaveDopplerPerformance}, should be advantageous because SVD filters provide more effective clutter suppression when they are employed with longer ensembles. Furthermore, the format of an angle-ordered spread-spectrum signal appears well suited to the higher-order angular-domain SVD filter introduced in \cite{jiang_clutter_2024}. Surprisingly, our initial two-dimensional imaging simulations of the spread-spectrum method in the presence of vibratory tissue motion suggest that a series combination of a notching comb filter with an SVD filter is more effective than an SVD filter alone when the tissue vibration is severe \cite{esmailian_background_2026}. Figure~\ref{fig:Fig16} provides a hint as to why this might be the case: eliminating the stationary components of the clutter signal may permit the clutter and blood signals to be more effectively separated among the remaining singular vectors. Further analysis of this topic using the spatial similarity matrix of the singular vectors \cite{baranger_adaptive_2018} is presented in \cite{esmailian_background_2026}.

Spread-spectrum Doppler can also be considered alongside anti-aliasing methods for ultrafast Doppler that function via modification of the transmit pulse sequence, e.g., \cite{posada_staggered_2016,podkowa_high-frame-rate_2018,jensen_estimation_2019,leroy_revisiting_2026}. Among these methods, the double-transmission scheme introduced in \cite{podkowa_high-frame-rate_2018} is an interesting comparator because that study also considered using low-resolution images acquired at multiple plane-wave angles for velocity estimation. In fact, the angle-ordered sequences investigated in Figs.~\ref{fig:Fig3}--\ref{fig:Fig5} and Sect.~\ref{subsec:CombFilterFrequencyEstimation} of this paper are equivalent to the single-transmission, uncompounded ``sT$\times$$M$ LRI'' sequences used as control examples in \cite{podkowa_high-frame-rate_2018}. For example, a three-angle, double-transmission sequence with a maximum $\alpha_{\mathrm{Tx}}=10^\circ$ would be $\{-10^\circ,-10^\circ,0^\circ,0^\circ,10^\circ,10^\circ\}$ repeated multiple times. An integer-periodic clutter signal is avoided in the double-transmission approach by evaluating the lag-one autocorrelation using only paired samples from consecutive transmissions at the same $\alpha_{\mathrm{Tx}}$.

Recently, the double-transmission method was merged with the staggered-PRF method from \cite{posada_staggered_2016} to obtain a method named ``StaBle'' \cite{wahyulaksana_stable_2026} that promises a factor of 6 to 12 increase in the Nyquist velocity. In the StaBle method, the multiple plane-wave angles are used to obtain slow-time signals at different Doppler angles for velocity vector imaging, as described in \cite{yiu_least-squares_2016}, so again no compounding is performed. The mouse-model demonstration in \cite{wahyulaksana_stable_2026} used a three-angle sequence with two staggered PRFs related by $\mathrm{PRF}_2=(2/3)\mathrm{PRF}_1$, where $\mathrm{PRF}_1$ is employed at $\alpha_{\mathrm{Tx}}=\pm 7.5^\circ$ and $\mathrm{PRF}_2$ is used at $\alpha_{\mathrm{Tx}}=0^\circ$. The paper states that use of $M=3$ angles yields a factor-of-three increase in the Nyquist frequency; this statement refers to restoring the Nyquist frequency from $\mathrm{PRF}_1/(2M)$ to $\mathrm{PRF}_1/2$, the same as the difference between the coherent-compounding and spread-spectrum methods in Table~\ref{tab:PlaneWaveDopplerPerformance}. The $2/3$ ratio between the PRFs yields an additional doubling of the Nyquist frequency \cite{posada_staggered_2016}, raising it to $\mathrm{PRF}_1$ in that example. As illustrated in Fig.~1(d) of \cite{wahyulaksana_stable_2026}, the time to perform one repetition of the StaBle pulse sequence to acquire one slow-time sample pair at each plane-wave angle is $5/\mathrm{PRF}_1 + 1/\mathrm{PRF}_2$. Forming a color Doppler image from seven repetitions of that StaBle sequence would yield a Doppler frame rate of $\mathrm{PRF}_1/45.5$, which is the closest possible frame rate to the $\mathrm{PRF}/45$ in Table~\ref{tab:PlaneWaveDopplerPerformance}. For a direct comparison to the spread-spectrum method, if the paired samples were used to estimate a single axial velocity as with the ``dT$\times$$M$ LRI'' double-transmission sequences in \cite{podkowa_high-frame-rate_2018}, velocity estimation would be performed with 14 sample pairs at $\mathrm{PRF}_1$ and 7 sample pairs at $\mathrm{PRF}_2$, which is equivalent to ensemble lengths of 15 and 8 samples, respectively, for sequences acquired sequentially at each PRF.

\section{Conclusions}
\label{sec:Conclusions}

For color Doppler applications where the background tissue can be assumed to be approximately stationary, the preferred implementation of spread-spectrum Doppler employs a frequency-sampling FIR notching comb filter applied via circular convolution. The comb filter functions as a clutter filter and, in combination with the pulse sequence randomization, also partially compensates for the lack of transmit focusing in the plane-wave low-resolution images. Further development of the spread-spectrum Doppler method is required to make it more robust to non-stationary clutter. There may also be opportunities to improve the method's performance by making better use of the information provided by imaging at multiple plane-wave angles without compounding, e.g., by introducing higher-order SVD clutter filters or by extending the method to velocity vector imaging.

The relative utility of spread-spectrum Doppler compared to other state-of-the-art methods for ultrafast imaging of high blood velocities ultimately depends upon the priority the system designer assigns to their flexibility in specifying the ensemble length for velocity estimation. Ensemble lengths greater than the 8--16 samples typically used in conventional focused-beam color Doppler reduce the variability of autocorrelation-based velocity estimates \cite{guidi_real-time_2021,loupas_axial_1995} and open the possibilities of improving clutter filtering performance via the use of SVD filters \cite{demene_spatiotemporal_2015} and computing pulsed-wave-like Doppler spectra for each pixel in the image \cite{bercoff_ultrafast_2011}. Single-plane-wave Doppler and spread-spectrum Doppler are the methods that maximize the ensemble length at a given frame rate. If the system designer prioritizes maximizing both the ensemble length and the Doppler frame rate, the simulations presented in this paper suggest that spread-spectrum Doppler is an attractive option because of its superior spatial localization of the blood signal compared to SPW Doppler.

\section*{Acknowledgements}

This research was funded by NSERC Discovery Grant 2026-06162 and by an internal Strategic Success grant from Western University.

\bibliographystyle{IEEEtranN}  
\bibliography{Beaonmodelreferences}

\end{document}